\documentclass[10pt,twocolumn]{article}
\RequirePackage{calc}
\usepackage{booktabs}
\usepackage{adjustbox}
\usepackage{graphicx}
\usepackage{amssymb}
\usepackage{url,hyperref}
\usepackage{multirow}
\usepackage{comment}
 
\providecommand{\keywords}[1]{\textbf{\textit{Keywords---}} #1}

\begin{document}

\title{Evaluating Feedback Strategies for Virtual Human Trainers}





\author{Xiumin Shang\\
 University of California, Merced\\
\texttt{xshang@ucmerced.edu}

\and

Ahmed Sabbir Arif\\
 University of California, Merced\\
 \texttt{asarif@ucmerced.edu}\\
 
\and
Marcelo Kallmann\\
 University of California, Merced\\
 \texttt{mkallmann@ucmerced.edu}

 }

\maketitle
\begin{figure*}[!htb]
  \centering
  \includegraphics[width=0.30\textwidth]{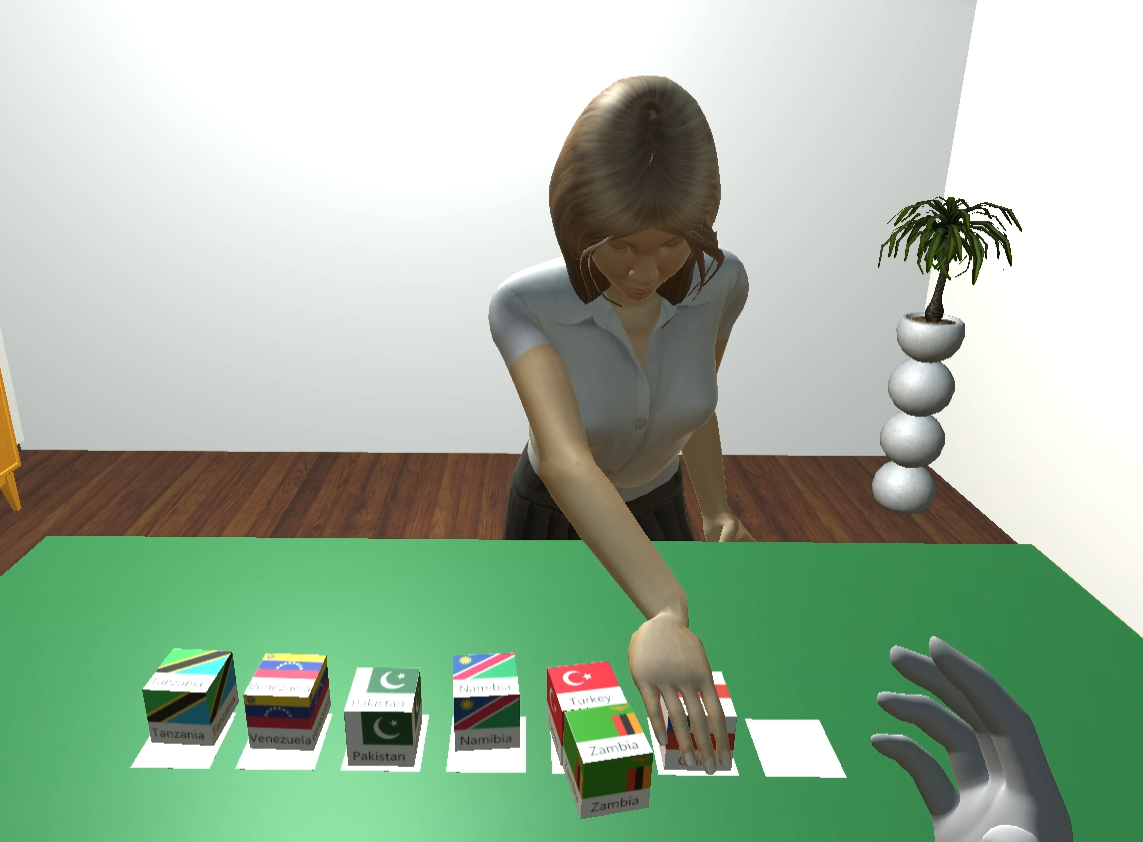}
  \hspace{0.2mm}
  \includegraphics[width=0.31\textwidth]{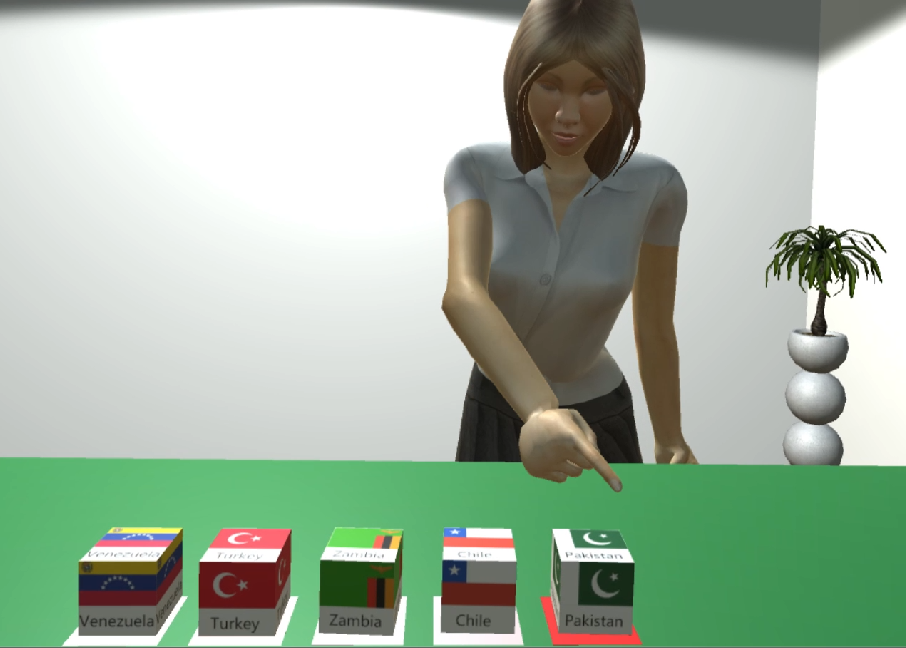}
  \hspace{0.2mm}
  \includegraphics[width=0.325\textwidth]{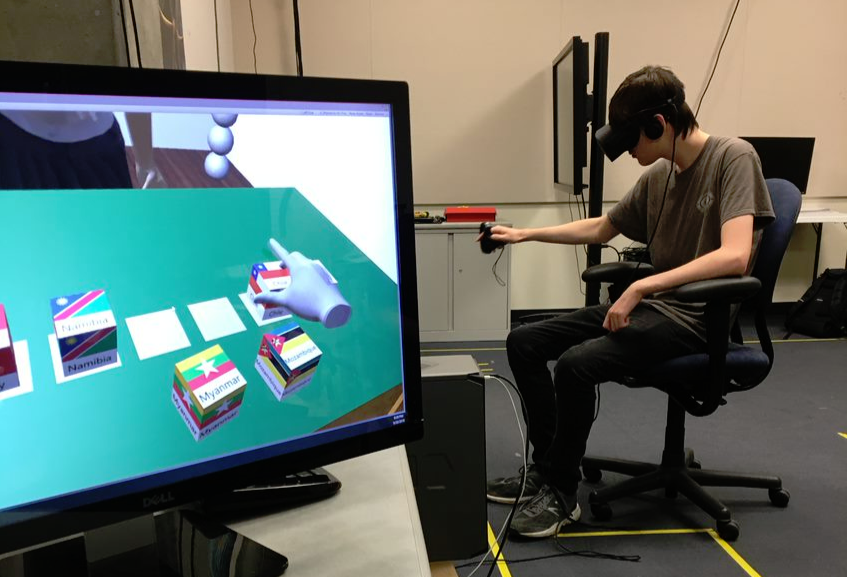}
  \caption{We evaluate an autonomous virtual trainer assisting users to perform a given task using different feedback strategies involving manipulation (left) and pointing (center).
  Users intuitively interact with the virtual trainer with voice commands while manipulating virtual cubes using Rift controllers (right).}
  \label{fig:teaser}
\end{figure*}

\begin{abstract}
In this paper we address feedback strategies for an autonomous virtual trainer. First, a pilot study was conducted to identify and specify feedback strategies for assisting participants in performing a given task. The task involved sorting virtual cubes according to areas of countries displayed on them.
Two feedback strategies were specified. The first provides correctness feedback by fully correcting user responses at each stage of the task, and the second provides suggestive feedback by only notifying if and how a response can be corrected. Both strategies were implemented in a virtual training system and empirically evaluated. The correctness feedback strategy was preferred by the participants, was more effective time-wise, and was more effective in improving task performance skills. The overall system was also rated comparable to hypothetically performing the same task with real interactions.
\end{abstract}



 






\section{Introduction}\label{sec:intro}

In this paper we identify, model, and evaluate feedback behaviors for virtual trainers assisting users to accomplish a given task involving manipulation of objects in a virtual world.
With the goal of minimizing task-specific characteristics, a simple sorting task was chosen based on sorting areas of countries represented on cubes. This task is composed of basic elements that are often present in a variety of scenarios: object manipulation for task execution, observation of results, and repetition in progressively difficult cases.

Given a task to be solved, we focus on the specification and evaluation of feedback behaviors for the virtual trainer, such that it can effectively demonstrate the task to be executed and provide feedback to assist the user to perform the task.
A pilot study was first conducted with a human trainer in order to identify and specify feedback strategies.
As a result of this study two feedback strategies were specified.
In the first feedback strategy the virtual trainer provides Correctness Feedback (CF) by fully correcting the responses of the users at each stage of the sorting task.
In the second feedback strategy the virtual trainer instead provides Suggestive Feedback (SF) by incrementally notifying the users if and how a current response is wrong in order to enable users to correct their responses by themselves.

An immersive VR system was then implemented to evaluate the two feedback behaviors. The system immerses the user with an Oculus Rift Head-Mounted Display (HMD), incorporates speech recognition and synthesis for communication, and implements a number of behaviors for replicating intuitive human-like interactions. See Figures~\ref{fig:teaser} and ~\ref{fig:design}.

Both CF and SF strategies represent valid approaches for a virtual trainer to follow. While CF explicitly presents corrected results at each stage of the task, the SF strategy requires the user to be engaged in correcting results by themselves. 
To our knowledge this work is the first to investigate such strategies in the context of direct interaction with a human-like virtual trainer in an immersive 3D VR environment.


We have collected performance data from  14 participants using our system employing  both feedback strategies. 
The CF strategy was preferred by participants, was more effective time-wise, and was more effective in improving task performance skills. 
The overall system was rated comparable to hypothetically performing the same task with real interactions. Additional findings include 
interaction with speech commands rated unfavorably, and HMD visual resolution and quality rated as not interfering or distracting from the task execution. 
A poster abstract was previously published about this project\cite{shang2019effects} and this paper presents our full work.



\section{Related Work}

Virtual environments have been employed in a number of applications in education, training and beyond~\cite{Dede2009,Freina2015APerspectives,Biljanovic2010IntelligentReview,Abulrub2011VirtualLearning, Xiumin2019}; and
the inclusion of autonomous virtual humans in such types of applications is a natural way to achieve effective human-like interactions in  virtual environments.

In order to be effective a number of specific behaviors have to be implemented in an autonomous virtual trainer. 
In this paper we address the specification, implementation and evaluation  of  feedback behaviors, and we also present the complete integration of the evaluated behaviors in our virtual human training system.



\subsection{Virtual Agents and Training Systems}
Virtual agents can help users to learn by utilizing a variety of behaviors based on gestures, natural language, gaze, and facial expressions~\cite{Lester1997TheAgents,Kirk2006ComparingTasks,07-Chaminade,Baylor2004PedagogicalRole,Xiumin2019,Smith2018CommunicationReality}. Well-designed non-verbal behaviors for virtual agents are in particular important as they can increase the user's attentiveness, positivity, and also rapport~\cite{Tickle-Degnen1990TheCorrelates}, which defines the ability to maintain harmonious relationships based on affinity~\cite{Granitz2009NowStudents}.

A number of training systems relying on virtual characters acting as interactive demonstrators, virtual teammates, virtual teachers and other roles have been developed. Steve~\cite{Rickel1997} represents one of the first systems implementing an autonomous virtual trainer specifically designed to train people to operate ship engines. Another example, among many others, is AutoTutor~\cite{Graesser2004} which is a virtual tutor designed to teach concepts in science and mathematics with multiple design strategies like using dialogue, feedback, corrective statements, hints, fill-in-the-blank questions, and requests for more information from the user. 
Significant past research has been dedicated to developing the necessary movement synthesis and behavioral modeling algorithms for accomplishing autonomous virtual humans and a number of software solutions have been developed for facilitating their integration in applications, such as the Virtual Agent  Interaction Framework (VAIF)~\cite{Gris2018VirtualAgents} and the Virtual Human Toolkit~\cite{Hartholt2013AllToolkit}.

\subsection{Feedback Strategies}


Several works have evaluated different types of feedback strategies in the context of classroom learning~\cite{Guenette2007IsWriting,Nelson2009ThePerformance,Shute2008FocusFeedback}. Shute~\cite{Shute2008FocusFeedback} provides an elaborate review of feedback types, organizing them as: no feedback, verification, correction, try again, error flagging, elaborate, attribute isolation, topic contingent, response contingent, hints/cues/prompts, bugs/misconceptions, and informative tutoring. Informative tutoring is the most complext type of feedback
which was defined as ``information communicated to the learner that is intended to modify his or her thinking or behavior to improve learning''. It represents
and it includes verification feedback, error flagging, and strategic hints. 
Some studies~\cite{bangert1991effects,Pridemore1995ControlInstruction} have reported that giving learners informative feedback was more effective than only giving them correct answers directly.

Previous works have also investigated feedback strategies in computer-based learning systems. Attali~\cite{Attali2015EffectsProblems} has tested four feedback types: no feedback, immediate knowledge of the correct response, multiple-try feedback with knowledge of the correct response, and multiple-try feedback with hints after an initial incorrect response, which was found to be the most effective type of feedback. Another study~\cite{VanderKleij2015EffectsOutcomes} reported that elaborated feedback during learning was superior to only providing answer correctness. Falcao~\cite{falcao2018FeedbackTabletop} combined interactive tangible tabletops 
and physical actions to study the discovery-based hands-on learning among children with intellectual disabilities. 
These works however have mostly studied feedback strategies in scenarios involving complex rules, complex knowledge retention, or physically involved tasks. For simple memory-based tasks  direct display of correct solutions has been identified as an effective feedback \cite{Shute2008FocusFeedback}.

While these previous studies provide several guidelines that can be used for implementing feedback behaviors,
 no previous work has directly investigated this topic in the context of a virtual trainer interacting with users in an immersive VR-based system.

One particular type of gesture  that is useful for delivering feedback behaviors is pointing.  Several works have investigated  pointing gestures for virtual humans to become able to identify spatial positions, a capability which is very important and which has been explored in several previous works~\cite{Huang2016,Rickel1997}. However, the use of pointing in feedback behaviors has  not been investigated. In our work we focus on interactive tasks where the user is required to manipulate virtual objects and the virtual trainer employs pointing as one way to deliver feedback during task execution.
 

In this paper we identify, specify and evaluate two feedback strategies of broad applicability: Correctness Feedback (CF), which focuses on only confirming correct answers, and Suggestive Feedback (SF), which provides informative feedback during the learning activity in our virtual training system.
We also present results which favor the use of the CF strategy.


\section{Pilot Study}

In order to specify effective feedback strategies we have conducted a pilot study with initial versions of our training scenario between two persons without the use of any computer assistance.

\subsection{Apparatus}

No computer devices were used during the pilot study. A real human played the role of the virtual trainer. Printed paper cards were used to represent information to be sorted. The material sets were printed on a square paper of 7 cm side, and material sets used in this study were not repeated in the final study.
The sorting task was based on sorting the information illustrated in the material sets.

\subsection{Design}
In the considered sorting task the users needed to memorize a limited amount of information. The entire task was executed in about 10 to 20 minutes. As a tutoring system, the goal was to design a simple scenario but which also offered  some challenge~\cite{Sklar2003AgentsThing}. 

\subsubsection{Memorial Materials and Questionnaires}
\begin{figure}[!htb]
     \centering
     \includegraphics[width=.38\linewidth]{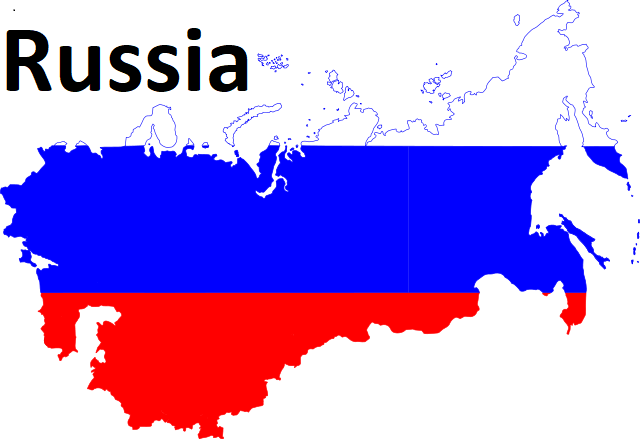}
     \hspace{0.6cm}
     \includegraphics[width=.38\linewidth]{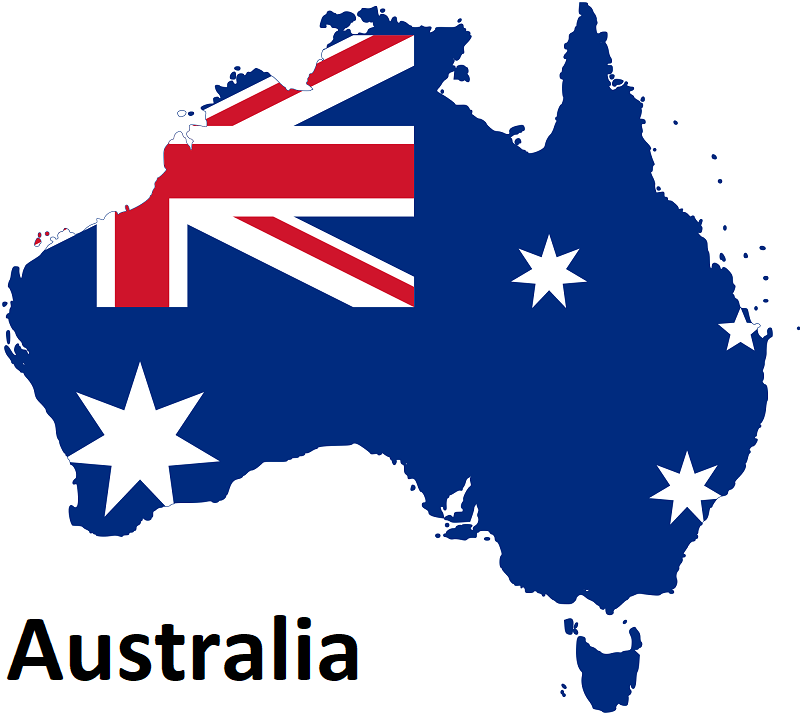}\\
\vspace{0.1cm}
     \includegraphics[width=.38\linewidth]{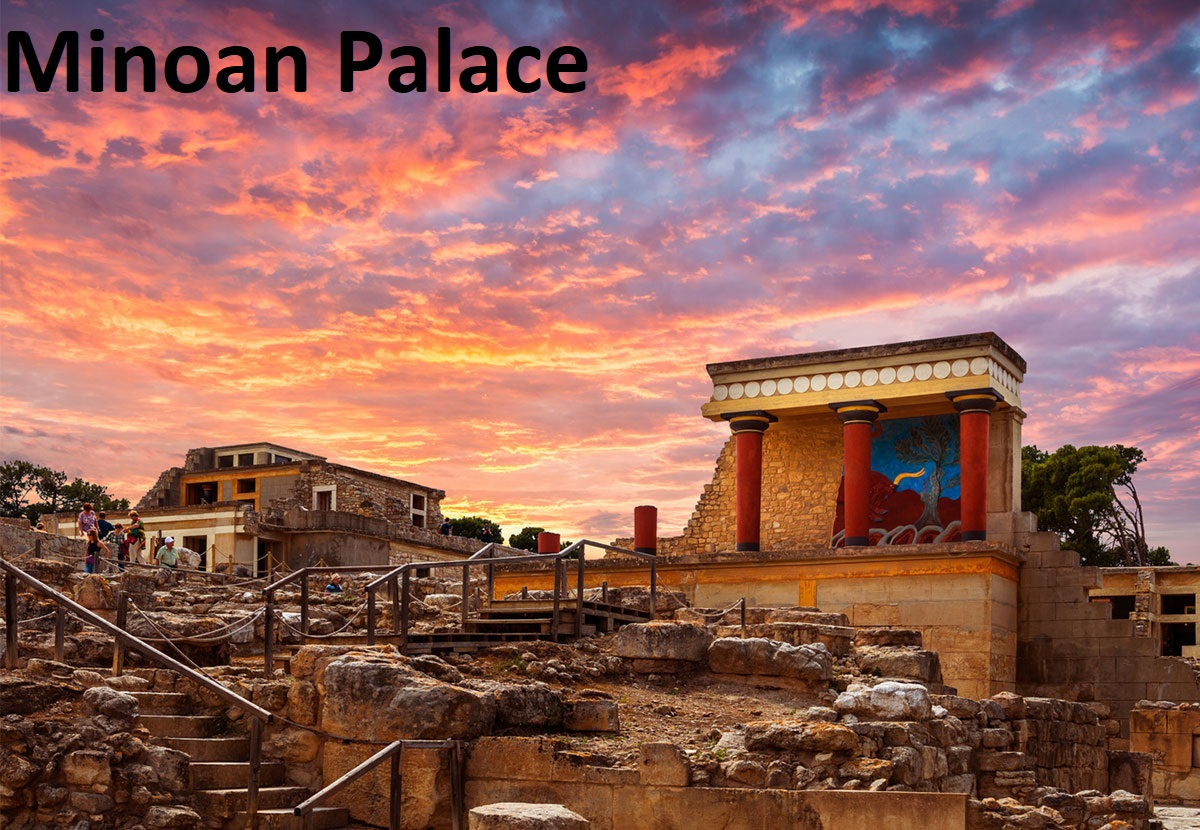}
     \hspace{0.6cm}
     \includegraphics[width=.38\linewidth]{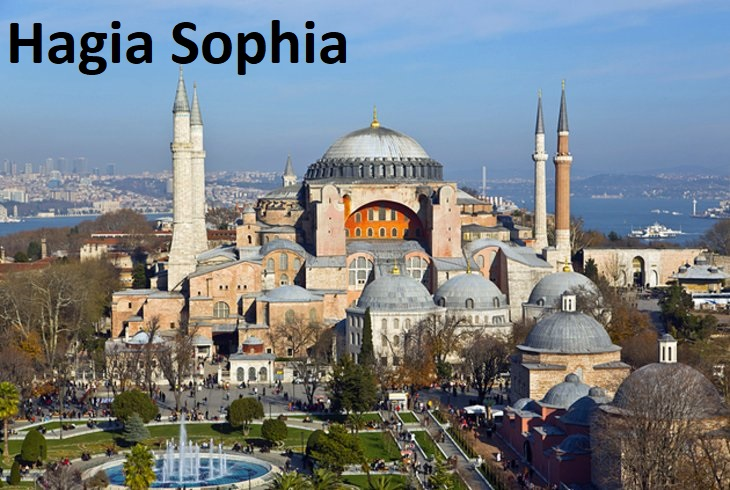}
   \caption{Example pictures used in the pilot study.}
   \label{Fig:pilotpics}
\end{figure}

Two different material sets were used during the study, each material set included a group of 9 pictures. The first set integrated the display of country maps, flags, and names, as shown in Figure~\ref{Fig:pilotpics}-top. The second set displayed ancient buildings and their names, as shown in Figure~\ref{Fig:pilotpics}-bottom.
The first material set had to be sorted from largest country land area to smallest country land area.
The second material set had to be sorted from oldest building to newest building.
In both cases the ordering was from the user's left side to the right side.
To quantitatively evaluate the user's performance we have applied pre- and post-test questionnaires. 
The pre-test questionnaire had one material set randomly printed, and the user needed to sort it based on his/her own previous knowledge. 
The post-test questionnaire had a new random material set, and was sorted by the user after the interactive activity described below. 

\subsubsection{Task Design}

Each sorting task consisted of sorting the cards gradually through interaction with the human trainer. Each sorting activity was organized in 4 stages. In stage 1, three cards were presented to the user on the table in random order. The user would pick the cards and sort them according to the sorting criterion. The human trainer then provided some sort of feedback until the three cards were sorted correctly. At this point, the task would proceed to the next stage where the same cards of the previous stage would appear again, but with two additional cards. In this way the user was supposed to incorporate the information on the two new cards when sorting the entire set of cards. 
Additional pairs of cards were added at each new stage until reaching stage 4, when nine cards were sorted by the user. 

\subsubsection{Study Design}
First, the members of our research team have
executed our pilot study scenario and, as discussed in Section~\ref{sec:intro}, two overall feedback strategies were identified: Correctness Feedback (CF) was based on correcting responses at each sorting stage, and Suggestive Feedback (SF) was based on just notifying,  at each stage, the cards that were sorted incorrectly. 

The pilot study then evaluated these strategies using two independent variables:
\begin{itemize}
\item Feedback. The two conditions of the independent variable ``feedback'' were: correctness feedback and suggestive feedback.
\item Stage.
The four conditions of the independent variable ``stage'' were: stage 1, stage 2, stage 3, and stage 4.
\end{itemize}

The two dependent variables used in this study were:
\begin{itemize}
\item Performance Improvement.
The pre-test and post-test questionnaires were used to record the users' sorting scores of the material sets, and the task performance improvement of a  participant was represented by his or her sorting scores.
\item Stage execution time.
The time duration taken for producing a sorting for each stage of the task. This time excludes any interaction time between the user and trainer.
\end{itemize}

A within-subjects design method was used. Each participant experienced the two feedback conditions and each with the two material sets. The order of feedback conditions and material sets was counterbalanced to reduce their effect on the dependent variables. In summary, the within-subjects design was: 
 4 participants $\times $ 
2 conditions (with two material sets) $\times$ 
 4 task stages $\times$ 
3, 5, 7, 9 country cards $=$ 192 cards in total.

\subsection{Participants}
Four volunteers were involved in our pilot study: two males and two females, with ages ranging from 27 to 32 years old, and all were English speakers. One of our researchers played the role of the human trainer.
Each participant performed the activity twice, each time experiencing a different feedback strategy and a different material set.
The order of delivering the two strategies was counterbalanced among the participants. 

\subsection{Procedure}
The study was conducted in an empty room. The human trainer first collected the oral consent from the participant and explained the study. Then, for each material set and feedback strategy, the participant performed the pre-test questionnaire, the main sorting activity while receiving feedback from the human trainer, and the post-test questionnaire.


\begin{figure*}[htb]
  \centering
  \includegraphics[width=\textwidth]{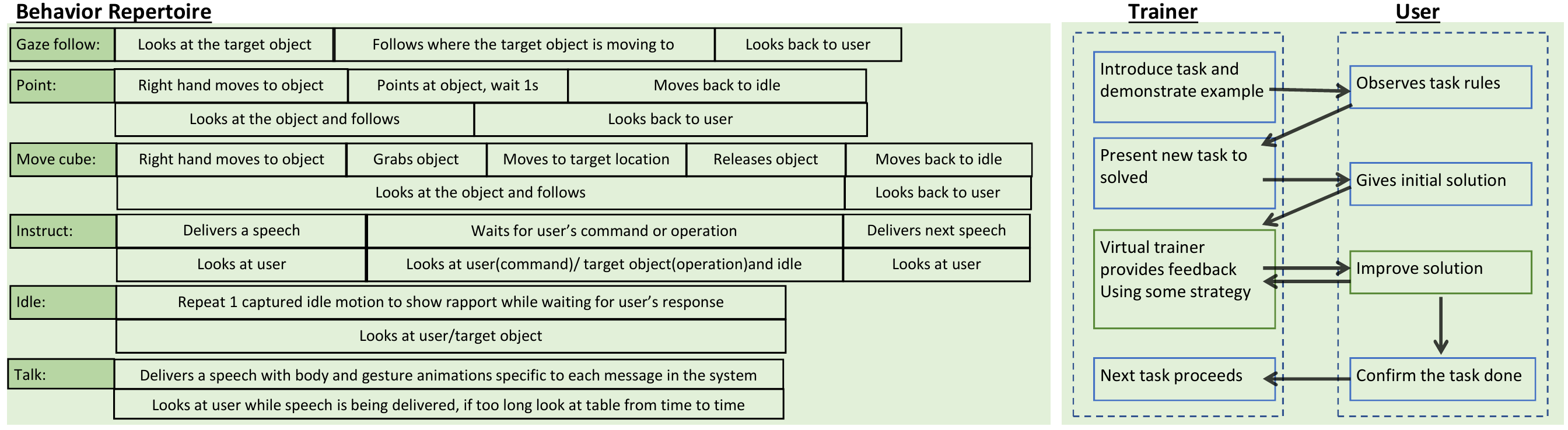}
  
  \caption{\label{fig:design}
  Left: the behavior repertoire available to the virtual trainer. Each behavior (except the first one) implements an action synchronizing arm motion and gaze movement as illustrated in the horizontal bars next to each behavior.
  Right: the main interaction phases between the virtual trainer and the user.}
\end{figure*}

With CF, when a wrong sorting was presented, the human trainer would correct the  sorting and say ``Here is the correct order, please study it and let me know when to continue''. With SF, the human trainer would point to each pair of cards incorrectly ordered and say ``These two cards are in the wrong order, please correct them''.
The whole study took around 30 minutes to finish.

\subsection{Results}

In order to evaluate the results from the pre-test and post-test questionnaires, we defined a ``sorting score'' to quantify a participant's performance improvement. For a set of 9 cards, the total possible combination of pairs is C(9,2)=36. The sorting score is defined as the number of pairs in correct order divided by the total number of pairs (36) and multiplied by 100 in order to express values as percentages.

With this definition, the mean scores of CF in the pre-tests and post-tests were 56.94\% and 95.83\% respectively. The mean scores of SF were 69.44\% and 99.31\% respectively. The mean time to complete the entire the task with the CF strategy was 229.8s, and with the SF strategy was 223.9s. These results show that both CF and SF were effective as they led to 68.30\% and 43.02\% performance improvement.
However, participants needed increasingly more time to complete each stage under SF than under CF, given that a growing number of suggestions were needed before completing each stage. 




All participants reported that the elements on the cards helped them memorize the needed information. For example, when the country material set was used, one participant said ``I can't remember the country's name, but I do remember the colors and shape of the country I was sorting''. Figure~\ref{Fig:pilotpics}-top illustrates the used shapes and colors. The land area of Russia is larger than Australia. In the building material set, one participant said ``It is hard to tell how old the building is, but those devastated buildings must be older than those completely built''.
Indeed, as is the case in Figure~\ref{Fig:pilotpics}-bottom, the Minoan Palace is older than Hagia Sophia.


In our main study we have changed the material sets and redesigned the questionnaires in order to better constrain the participant to assimilate the intended information. We have also updated the delivery of the SF strategy to reduce duration times and participant stress due to possible large number of suggestions. These changes are presented in Section~\ref{sec:study}.



\section{System Implementation}


A training system was implemented providing the needed capabilities for executing the required interactions observed in our pilot studies. The system was built with the Unity game engine connected to an Oculus Rift.  We use the Oculus Unity Integration plugin to connect to   the Rift controllers. The general system is designed as follows: the system monitors the user input, runs a task-dependent action decision model, and then decides which behavior the virtual trainer has to execute. This process is repeated until the simulated task ends. This interactive process is illustrated in Figure~\ref{fig:design}-right.

In order to be able to execute all needed actions, the virtual trainer was equipped with six basic behaviors, as detailed in Figure~\ref{fig:design}-left.
The movement of the ``gaze follow'' behavior is controlled by Inverse Kinematics (IK); for ``point'' and ``move cube'' the right arm is also controlled by IK while the hand shape is interpolated between a neutral hand shape and the target hand shape (pointing or grasping shape). The ``talk behavior'' includes  lip syncing implemented with the Unity plugin SALSA. User voice commands are analyzed by the Windows Speech Recognition module. The ``idle'' behavior is  animated with motion-captured animations captured with a Perception Neuron full-body motion capture suit.
In order to be realistic most of the behaviors synchronize arm movements with a gaze attention model specifically designed for each behavior as depicted in the ``timeline boxes'' in Figure~\ref{fig:design}-left.
This behavior repertoire was sufficient to implement the target scenarios and feedback strategies.

\section{User Study}~\label{sec:study}


Our main user study took place at our research laboratory equipped with a high-end computer station connected to the Oculus Rift and running our simulation system.



\subsection{Design}

Our simulation system replicated the same overall scenario of the pilot study, but with several improvements.

\begin{figure}[!htb]
   \begin{minipage}{0.23\textwidth}
     \centering
     \includegraphics[width=.9\linewidth]{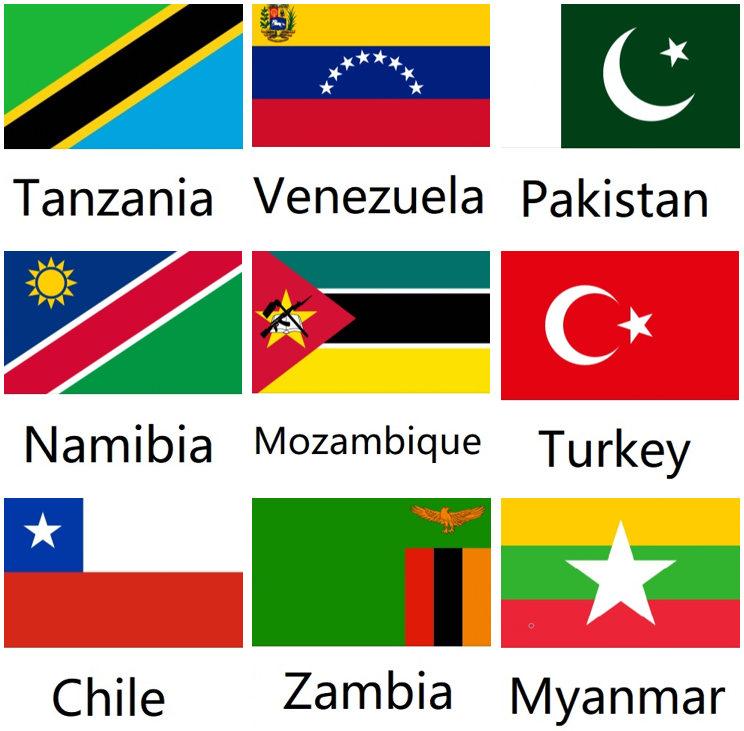}
   \end{minipage}\hfill
   \begin{minipage}{0.23\textwidth}
     \centering
     \includegraphics[width=.9\linewidth]{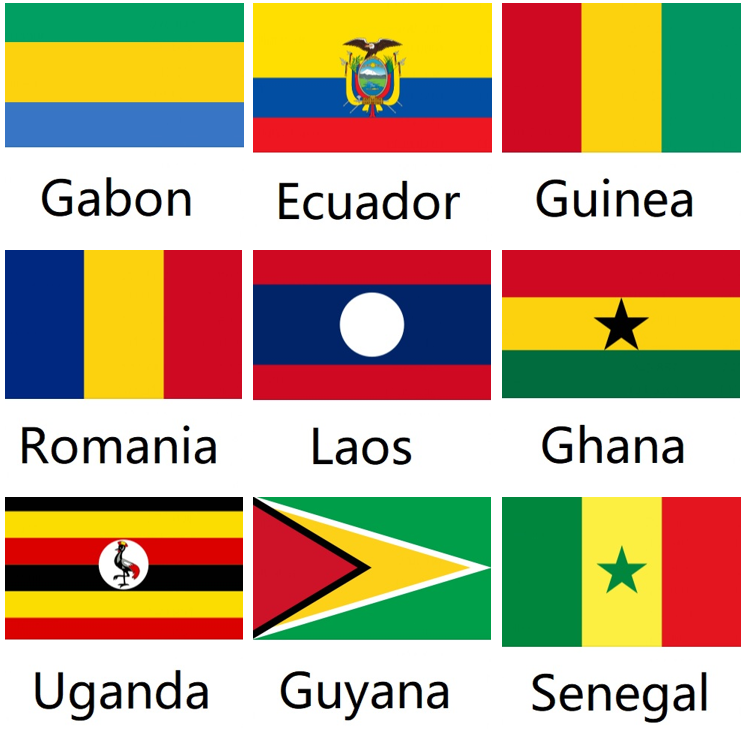}
   \end{minipage}
   \caption{Country material sets used during main study}
   \label{fig:picsets}
\end{figure}


\subsubsection{Memorial Materials and Questionnaires}
Only country materials were used.
We used one material set in order to evaluate users in a same topic and discarded the building material set because it allowed different types of information to influence the participants.
The country material set was also updated to reduce non-applicable information influence, and country border outlines were remove.
Two non-intersecting sets of countries were used (see Figure~\ref{fig:picsets}-left and Figure~\ref{fig:picsets}-right). 
The sets were chosen such that the countries have similar land areas. 
The pre-test and post-test questionnaires have been correspondingly updated to only list countries by names.

\subsubsection{Task Design}
The overall task procedure is the same as performed in the pilot study; however, the task now is based on sorting virtual cubes representing countries instead of sorting cards.



\subsubsection{Feedback Strategies}

\begin{figure*}[!htb]
  \centering
  \includegraphics[width=0.32\textwidth]{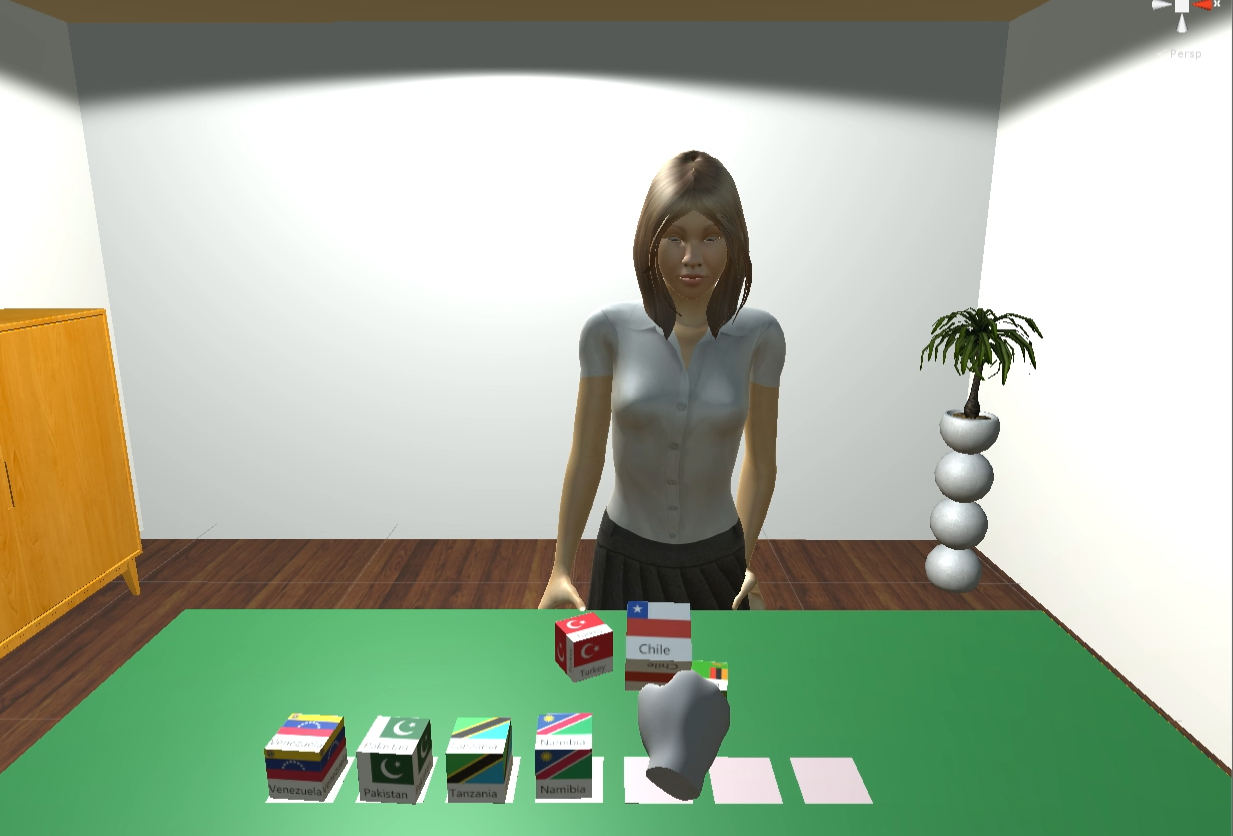} 
  \includegraphics[width=0.31\textwidth]{figures/cf22.png}
  \includegraphics[width=0.32\textwidth]{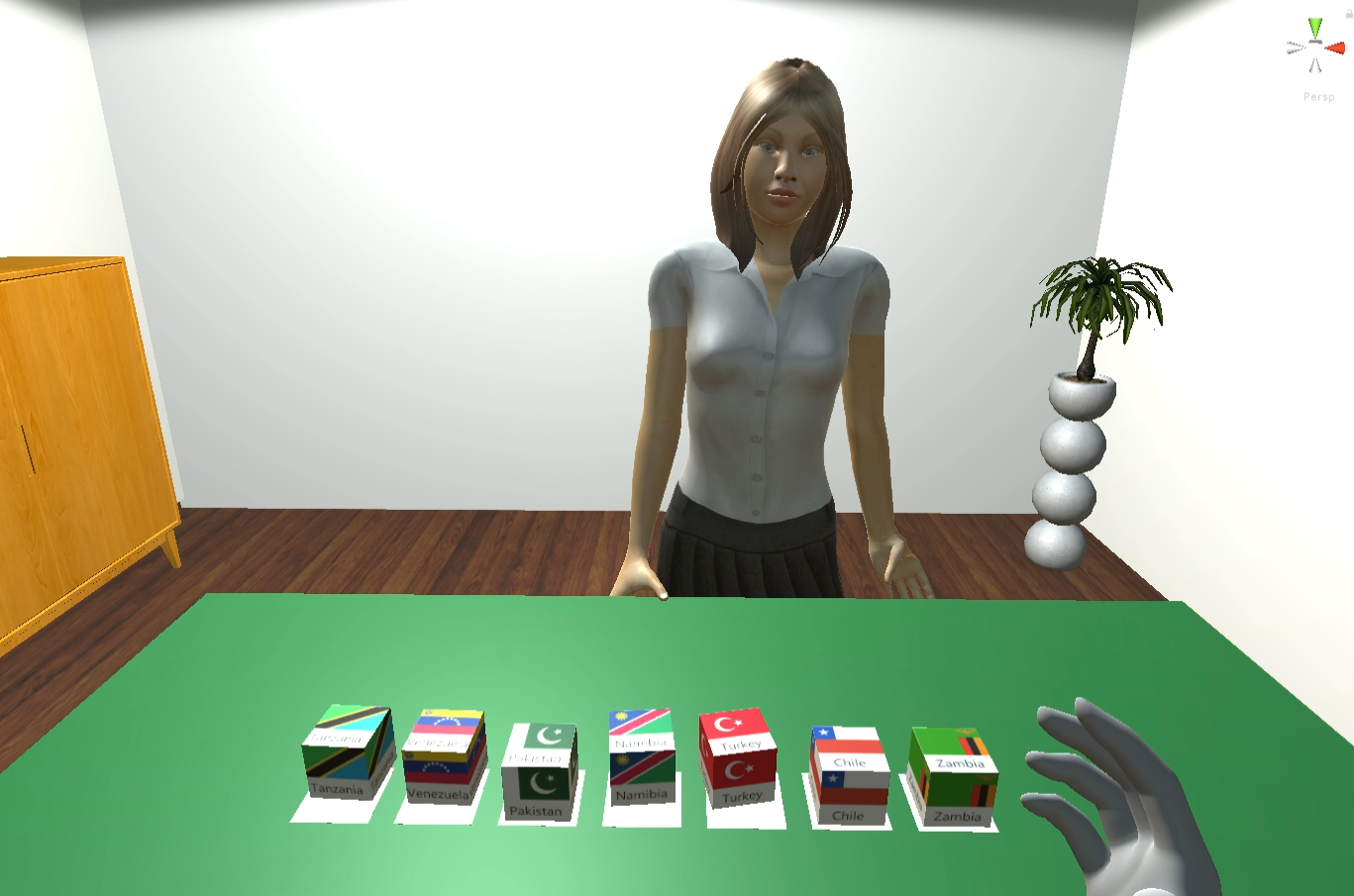}\\
  (a) \hspace{5.4cm} (b) \hspace{5.4cm} (c)\\
  \vspace{3mm}
  \includegraphics[width=0.32\textwidth]{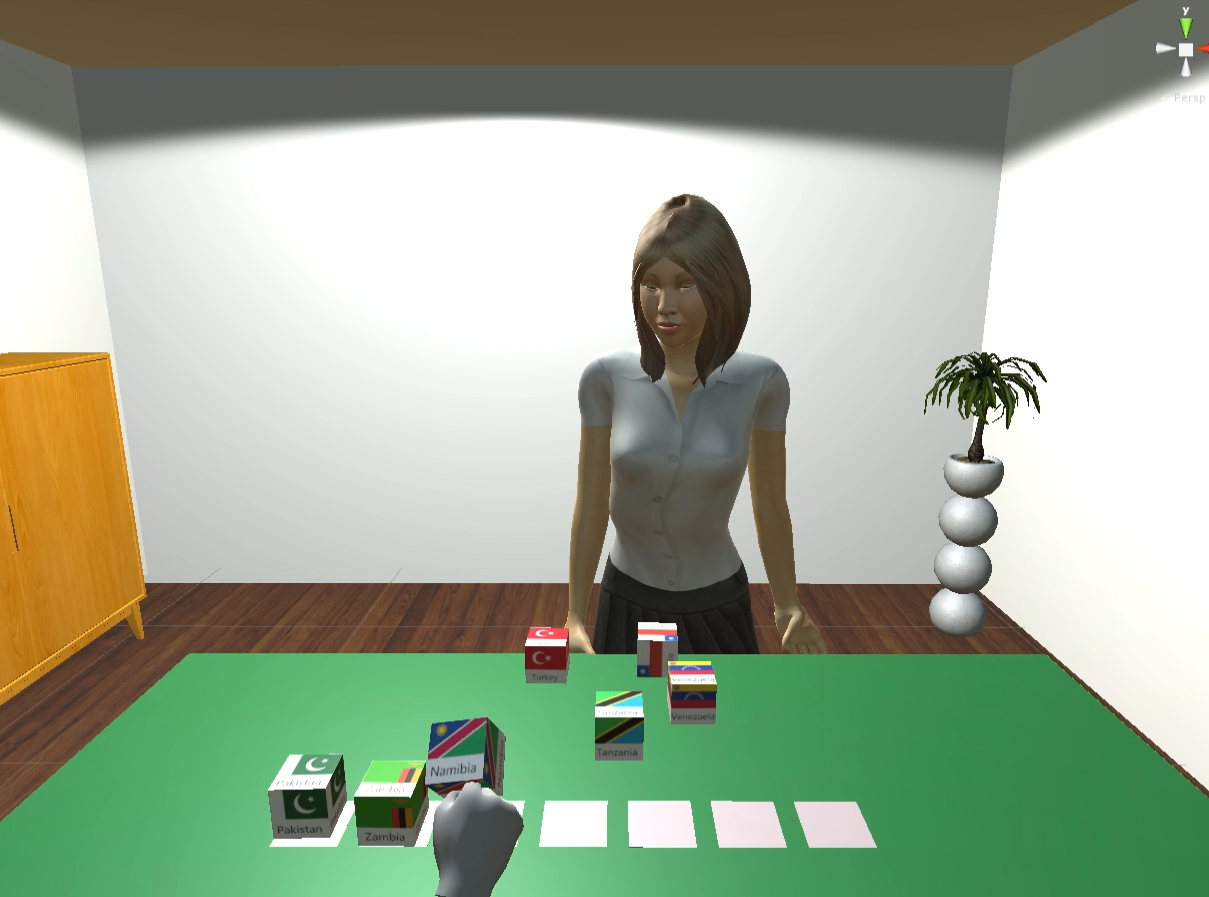} 
  \includegraphics[width=0.32\textwidth]{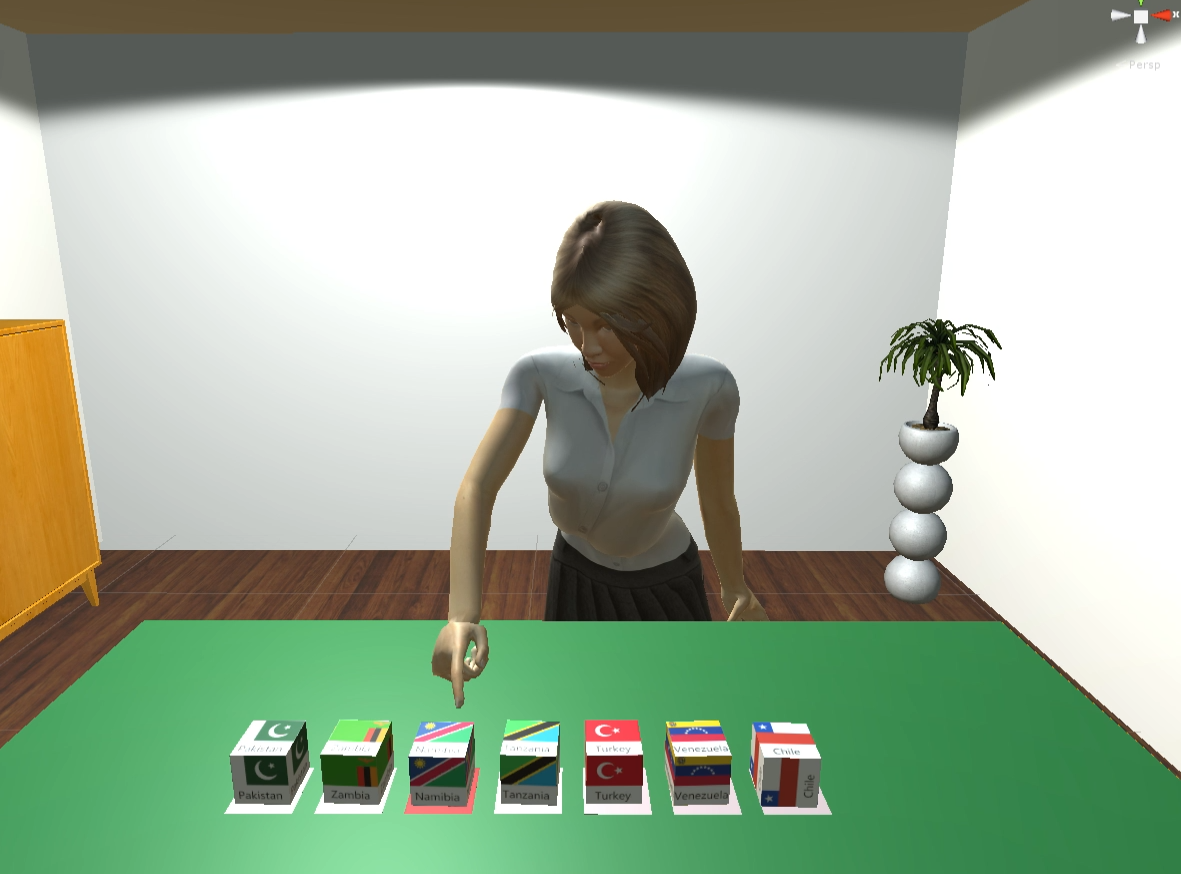}
  \includegraphics[width=0.325\textwidth]{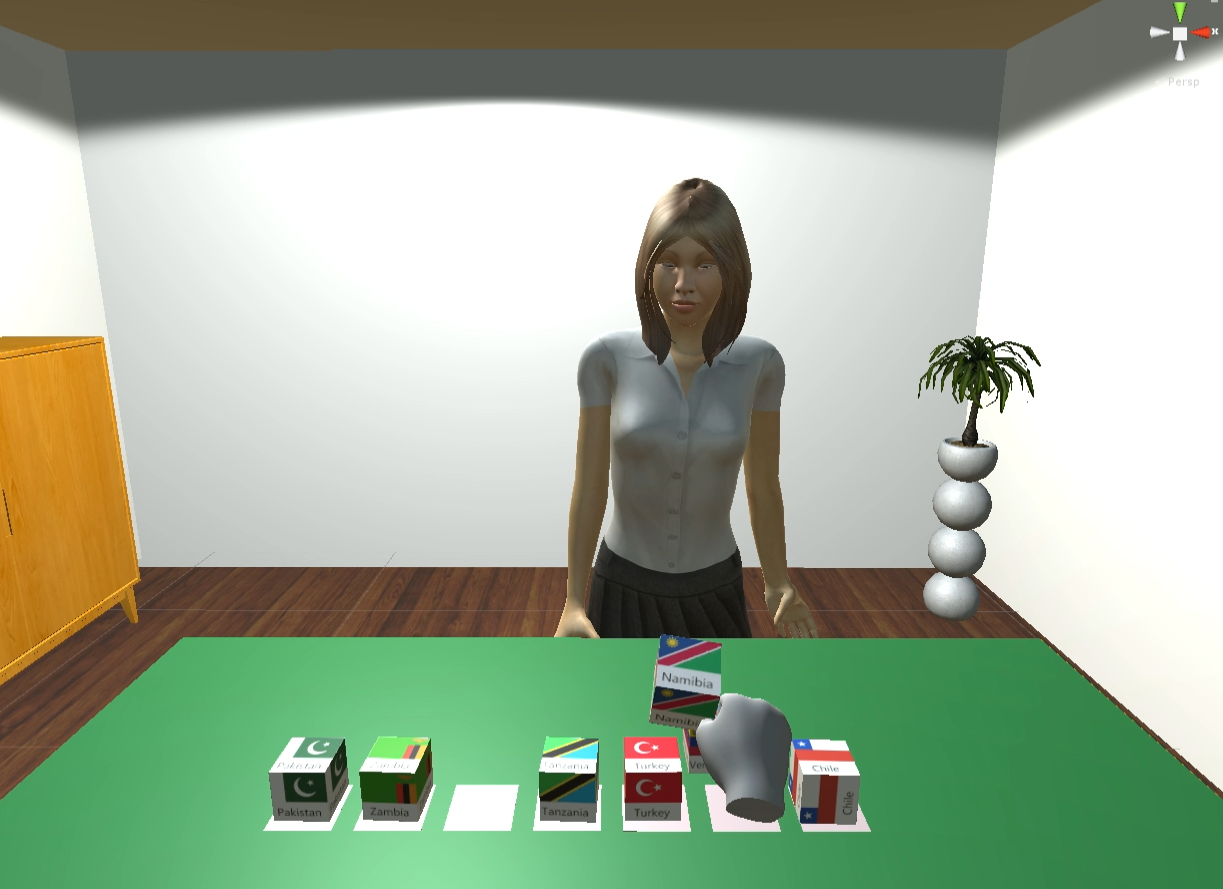}\\
  (d) \hspace{5.4cm} (e) \hspace{5.4cm} (f)\\

  \caption{\label{fig:strategies}
  The Correctness Feedback (CF) strategy is illustrated in images a-c. After the user arranges the cubes (a) and completes an incorrect sorting in the current task stage, the virtual trainer will manipulate the cubes and re-arrange them in the correct order (b). The user can then observe the correct solution as long as needed (c) until saying ``continue'' in order to move to the  next stage.
  The Suggestive Feedback (SF) strategy is illustrated in images d-f. After the user manipulates the cubes (d) and completes an incorrect sorting of the cubes in the current task stage the virtual correcter will then point to the cubes which are in the wrong position (e). The virtual trainer will point to all of the cubes that are wrong and say that their positions are wrong.  The user will then rearrange the cubes (f) to propose a new sorting. The process repeats until all cubes are placed in correct locations. When this is detected the virtual trainer notifies that the solution is correct and the user can then observe the correct solution as long as needed  until saying ``continue'' in order to move to the  next stage.
  }
\end{figure*}

Based on feedback obtained during the pilot study, the SF strategy was updated in the following way: instead of showing wrong pairs of items one at a time, the virtual trainer now points to all the wrong items at the same time. In this way, when there are more than 2 wrong cubes, the feedback is delivered in a shorter time and participants have to be more engaged since correcting the answer is not anymore obvious and mechanical.
When the user provided a wrong sorting the virtual trainer would say  ``Attention, those [two/three.../nine] countries are in the wrong order'' while pointing at those wrong countries one by one. 
In order to facilitate the perception of the users white squares marking the location of the cubes turned red each time the virtual trainer pointed at them.
Figure~\ref{fig:strategies} illustrates the execution of both strategies by the virtual trainer in our system.


\subsubsection{Study Design}
The same within-subjects design was used in the main study but with 14 participants: 
14 participants $\times$ 
2 conditions (with different learning material sets) $\times$ 
4 task stages $\times$ 
3, 5, 7, 9 country cubes=672 cubes in total.

\subsection{Participants}
14 volunteers from the university community participated in the user study.
None of them were involved in the pilot study. Their age ranged from 17 to 25 years old, with an average of 20.14 years (SD=1.99). They were all fluent in English, 5 of them were female and 9 male, 5 were left-handed and 9 right-handed.
Only 3 participants had experienced VR before, among them only 1 had used Oculus Rift before participating in our study. Participants received a \$5 cash as compensation for their time to our study.


\subsection{Procedure}

Before the test day participants received an email explaining  the purpose and main procedures of the study. Participants came in one by one at their scheduled times.
They were introduced again to our study, signed the consent form, and filled a demography questionnaire.
The user then completed two learning activities (one with SF and the other with CF) including the pre-test and post-test questionnaires.
Two additional post-study questionnaires were included: a simulator sickness questionnaire (SSQ) and a questionnaire asking users to rate the feedback strategies, the virtual trainer character, and the overall VR experience with 7-point Likert scales.
The system recorded all completion times for later analysis. Overall, each participant spent around 50 minutes to go through all activities. 
Figure~\ref{fig:teaser}-right illustrates one participant in the study.




\section{Results}

We summarize in this section our results.

\subsection{Qualitative Results}


The Simulator Sickness Questionnaire (SSQ) obtained the average total simulated sickness score of 9.82 (SD=9.8); the average nausea score of 19.08 (SD=18.78), with 11 participants reporting this symptom; the average oculomotor score of 3.79 (SD=5.99), with 6 participants reporting this symptom; and the average disorientation score of 1.75 (SD=4.62), with only 2 participants reporting this symptom. 

The Simulator Sickness Questionnaire (SSQ) analyzed scale means is shown in Table~\ref{tab:ssq} (in Appendix). The average total simulated sickness score for the system is 9.82 (SD=9.8). The average nausea score is 19.08 (SD=18.78), 11 participants reported this symptom; the average oculomotor score is 3.79 (SD=5.99), 6 participants reported this symptom; and the average disorientation score is 1.75 (SD=4.62), only 2 participants reported this symptom. 

\subsection{Quantitative Results}


We used a repeated-measures ANOVA with alpha of 0.05 for all analyses. ANOVA failed to identify a significant effect of feedback on performance improvement (\textsl F$_{1,13}$=1.66, \textsl p=.22). 
ANOVA failed to identify the significant impact of feedback and pre-test and post-test questionnaires test order on the user's performance improvement, the result is:\textsl F$_{1,13}$=0.12, \textsl p=.73. ANOVA identified a significant impact of pre-test and post-test questionnaires test order on user's performance improvement, the result is:  F$_{1,13}$=49.44, \textsl p=0.000009. The sorting scores representing the performance improvement of all 14 participants are displayed in Table~\ref{tab:sortingscore1}.

\begin{table}[!h]
  \centering
  \begin{adjustbox}{width=\columnwidth}
  \begin{tabular}{l c c c c}
      {}
      &\multicolumn{2}{c}{\small{\textbf{CF Sorting Score (\%)}}}
      &\multicolumn{2}{c} {\small{\textbf{SF Sorting Score (\%)}} }\\
    &{\textit{pre-test}}
    & {\textit{post-test}}
      & {\textit{pre-test}}
    & {\textit{post-test} }\\
     \midrule

Mean	&	44.84	&	77.78	&	51.19	&	81.15	\\
SD	&	14.16	&	20.20	&	12.19	&	19.05	\\

     \midrule
  \end{tabular}
  \end{adjustbox}
  \caption{Mean and standard deviation for obtained scores.}~\label{tab:sortingscore1}
\end{table}

\begin{figure}[th]
  \centering
  \includegraphics[width=0.98\columnwidth]{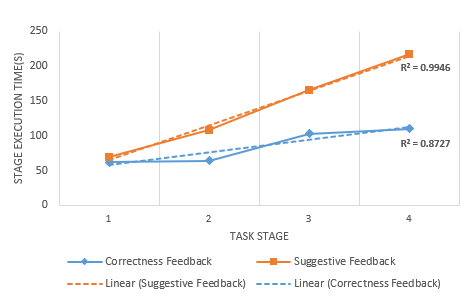} 
  \caption{\label{fig:logdata}
  Average stage execution times with CF and SF.
           }
\end{figure}

The average execution time for each stage of the task has been computed and the result is shown in Figure~\ref{fig:logdata}. 
The average total time needed to complete the tasks with CF was 339s, while with SF was 559s, which is 65\% higher than the CF time.



Two  linear regressions were calculated to predict the average stage execution time based on stage under both feedback strategies. For CF,  no significant regression equation exists (F(1,2)=13.71, \textsl p$>$.05). But for SF, a significant regression equation was found (F(1,2)=365.36, \textsl p$<$.05) with $R^{2}$ of .995. 



    


\subsection{Questionnaire Analysis}

In addition, our post-study questionnaire provided 14 questions to evaluate several aspects of the system.
The Wilcoxon Signed-Rank test was conducted for questions 1-5 but failed to identify a significant impact of feedback on participant's choices. However, for question 1,2 and 4, the median values of CF are higher than SF. Questions 6-14 are single questions about participants opinions on the system, virtual trainer, and the overall VR experience. The results are listed in Table~\ref{tab:poststudy}.

\begin{table*}[!ht]

  \begin{adjustbox}{width=\textwidth}
    \begin{tabular}{l l c c c c c c}

    {\textbf{ \#}}
    &{\textbf{Question}}
    &{\textbf{\textsl z}}
    & {\textbf{\textsl p}}
      & {\textbf{M$_C$$_F$}}
    & {\textbf{M$_\textit{S}$$_\textit{F}$} }
    &{\textbf{Md}}
    &{\textbf{Mn}}\\

    \midrule
    1	&	I think feedback x was effective for learning the given task	&	-0.63	&	.53	&	6	&	5	&	-	&	-	\\
2	&	I would imagine that most people would like to use feedback x for learning the task	&	-0.47	&	.64	&	5.5	&	5	&	-	&	-	\\
3	&	Feedback x made me feel bored/tired sometimes	&	-0.51	&	.61	&	3	&	3	&	-	&	-	\\
4	&	
I prefer to use feedback x for learning instead of
learning from a real human	&	-0.45	&	.65	&	4.5	&	3.5	&	-	&	-	\\
5	&	I think having the animated character in feedback x did not help to complete the task	&	-0.05	&	.96	&	3.5	&	3.5	&	-	&	-	\\
6	&	I like to use speech command with both feedback.	&	-	&	-	&	-	&	-	&	4	&	4	\\
7	&	I prefer using button command than speech command in the system	&	-	&	-	&	-	&	-	&	5.5	&	5.21	\\
8	&	I liked the way the character looked at me when it talked with me	&	-	&	-	&	-	&	-	&	4	&	4.21	\\
9	&	It would be important to see more human-like behaviors from the virtual character	&	-	&	-	&	-	&	-	&	5	&	4.57	\\
10	&	I would like to interact more with the VT in order to complete the tasks in the VE	&	-	&	-	&	-	&	-	&	4	&	4.14	\\
11	&	I felt comfortable/relaxed while doing the tasks in the virtual environment	&	-	&	-	&	-	&
-	&	6	&	5.5	\\
12	&	The visual display quality interfered or distracted me from performing the activities	&	-	&	-	&	-	&	-	&	2.5	&	2.86	\\
13	&   I was more proficient interacting with the VE at the end of the experience than start	&	-	&	-	&	-	&	-	&	5.5	&	5.64	\\
14	&	I felt completely immersed/involved in the VE during the execution of the tasks	&	-	&	-	&	-	&	-	&	6	&	5.64	\\
    \midrule

    \end{tabular}
  \end{adjustbox}
\caption{Post-study questionnaire Wilcoxon Signed-Rank test result and  median (Md) \& mean (Mn) values. VT stands for virtual trainer, VE stands for virtual environment.}~\label{tab:poststudy}

\end{table*}

\subsection{Discussion}


Some participants reported slight discomfort for experimenting our system, but none of the symptoms exceeded nausea, oculomotor, or disorientation severity levels. 
The total SSQ score of the system was 9.82, which is negligible (the maximum score possible on the SSQ is 300). Hence the system was appropriate for the study since its side effects were not severe enough to impact user performance or preference.

For both  CF and SF strategies the mean post-test sorting scores were significantly higher compared to their respective mean pre-test sorting scores after experiencing the learning activity in our system.
Statistical tests failed to identify a significant difference of CF and SF strategies on user's performance outcome; however, the increments were 72.79\% for CF and 58.53\% for SF, showing that CF was more effective than SF for increasing the performance outcome.
A significant linear regression equation for stage execution time was found for SF, but not for CF, and the total average time needed to complete a task with SF was 65\% longer than with CF. The reason is because  numerous suggestions were given by the virtual trainer, in particular as the stage number increased.

Overall, although no significant difference has been identified for both strategies, CF is more effective in the mean score increment aspect. From the analysis of task execution time, CF was more efficient time-wise for our task scenario. 
These findings agree with the notion that ``complexity of feedback may be inversely related to both ability to correct errors and learning efficiency''~\cite{Kulhavy1985FeedbackEfficiency,Shute2008FocusFeedback}.

For the post-study questionnaire analysis (Table~\ref{tab:poststudy}), no significant impact of feedback strategies on user's choice has been identified. However, comparing the median values of CF and SF, more participants thought CF was more effective for performing the sorting task and more participants mentioned  preference to use CF for performing the sorting task. 
Besides, our annotations show that 5 participants were uncomfortable when the virtual trainer repeatedly pointed out they did wrong sortings in a given task stage, which made them to not proceed the task session. One of the participants demonstrated significant anxiety at the last stage of a SF session even though the participant insisted in finishing the task. No one reported negative comments about the CF strategy during the user study. One participant said ``The first test (the CF test) was much faster and easier for me to use''.
These points show that the CF strategy was preferable to the SF strategy in our system.

Interestingly we found out that the speech recognition model in the system was not as welcomed as we expected. In question 6 participants rated the speech command in the median and average values both at 4, which means slightly approving it. However in question 7, when asked about the option of button commands, participants answered they would prefer using button commands than speech commands.
This could be explained by some participants having different language backgrounds, but even if not, they all had to pronounce a standard English accent in order for the system to correctly understand commands, otherwise they had to repeat voice commands several times until recognition in order to proceed with the session. We noticed that after three times repeating the command ``continue'' without being recognized the participant's voice volume would become lower and lower, demonstrating frustration, and in this case the experimenter manually instructed the system to proceed with a keyboard key without the participant noticing it. Participants rated the gaze tracking behavior as 4, probably because most of their attention was focused on the task state on the virtual table. In any case most of them expected to see a more human-like virtual trainer than the one we presented in our system. 

Another interesting finding is that participants rated very low the possibility that the visual display quality interfered or distracted them from performing the required activities. The median value was only 2.5. Perhaps, given that only 3 of the 14 participants had previous experience with a VR system, participants felt more excited to experience the VR display than frustrated from the drawbacks of using it.
Participants rated highly (rating 6) that they felt comfortable, relaxed and immersed while performing the tasks in the virtual environment.

Question 8 evaluated the gaze behavior incorporated in our system and participants rated it as an effective behavior of the virtual trainer. The evaluations of questions 10, 11 and 14 indicate that the design of the virtual trainer in our system was well received and effective for conducting the learning tasks. The evaluation of question 13 also indicates user improvement after being immersed in our VR system. In these questions the mean and median ratings show positive ratings from users.

Overall
our results indicate that in a short-memory training task such as the one we have used, the virtual trainer would be better designed with the correctness feedback strategy.
This result matches well with Shute's conclusion in traditional teaching feedback strategies~\cite{Shute2008FocusFeedback}: that when the task is simple, memory-based, and non-physical, the best performance is obtained with simple feedback features: correct solution, computer-delivered, and goal setting.

A number of interesting directions are possible for future work. It will be interesting to investigate further variations of feedback strategies and as well to apply the strategies to other types of tasks involving more complex object manipulation where task completion involves more complex physical movements instead of only memorization of concepts or properties. 
Furthermore, adaptation of the feedback strategies to the progress achieved during long-term tasks is  an important area to be investigated.

\section{Conclusion}
This paper presents a VR training system featuring an autonomous virtual trainer able to interact with users and provide feedback during task execution.
The effectiveness of two feedback strategies (CF and SF) was evaluated.
Although no statistically significant difference has been identified between CF and SF
with respect to performance outcome, CF was quantitatively found to 
lead to higher overall scores used to measure task outcome.
CF also showed to be the superior design in terms of being more efficient time-wise and being the preferred strategy of the users.


Our findings also support that in general interactions with virtual trainers were rated by users as comparable in preference to hypothetically performing the same task with real interactions.
These results match with previous work suggesting the effectiveness of interactions with virtual humans.
Overall our evaluations indicate that our system was well designed and incorporates effective strategies for interaction with users. We believe our results can find several applications in a number of interactive scenarios and applications.





\bibliographystyle{plain}
\bibliography{reference}

\begin{thebibliography}{10}

\bibitem{Abulrub2011VirtualLearning}
Abdul-Hadi~G. Abulrub, Alex~N. Attridge, and Mark~A. Williams.
\newblock {Virtual reality in engineering education: The future of creative
  learning}.
\newblock In {\em 2011 IEEE Global Engineering Education Conference (EDUCON)},
  pages 751--757. IEEE, 4 2011.

\bibitem{Attali2015EffectsProblems}
Yigal Attali.
\newblock {Effects of multiple-try feedback and question type during
  mathematics problem solving on performance in similar problems}.
\newblock {\em Computers {\&} Education}, 86:260--267, 8 2015.

\bibitem{bangert1991effects}
Robert~L Bangert-Drowns, James~A Kulik, and Chen-Lin~C Kulik.
\newblock Effects of frequent classroom testing.
\newblock {\em The Journal of Educational Research}, 85(2):89--99, 1991.

\bibitem{Baylor2004PedagogicalRole}
Amy~L. Baylor and Yanghee Kim.
\newblock {Pedagogical Agent Design: The Impact of Agent Realism, Gender,
  Ethnicity, and Instructional Role}.
\newblock In {\em International Conference on Intelligent Tutoring Systems},
  pages 592--603. Springer, Berlin, Heidelberg, 2004.

\bibitem{Biljanovic2010IntelligentReview}
Petar. Biljanovic, Croatian~Society for Information, Electronics
  Communication~Technology, and Microelectronics-MIPRO.
\newblock {Intelligent Pedagogical Agents in immersive virtual learning
  environments: A review}.
\newblock In {\em MIPRO 2010 : 33rd International Convention on Information and
  Communication Technology, Electronics and Microelectronics}, Opatija,
  Croatia, 2010. Croatian Society for Information and Communication Technology,
  Electronics and Microelectronics.

\bibitem{07-Chaminade}
Thierry Chaminade, Jessica Hodgins, and Mitsuo Kawato.
\newblock Anthropomorphism influences perception of computer-animated
  characters's actions.
\newblock In {\em Social Cognitive and Affective Neuroscience}. Books (MIT
  Press, 2007.

\bibitem{Dede2009}
Chris Dede.
\newblock {Immersive interfaces for engagement and learning}.
\newblock {\em Science}, 323(5910):66--69, 2009.

\bibitem{falcao2018FeedbackTabletop}
Taciana~Pontual falc{\~{a}}o and Taciana Pontual.
\newblock {Feedback and Guidance to Support Children with Intellectual
  Disabilities in Discovery Learning with a Tangible Interactive Tabletop}.
\newblock {\em ACM Transactions on Accessible Computing}, 11(3):1--28, 9 2018.

\bibitem{Freina2015APerspectives}
Laura Freina and Michela Ott.
\newblock {A Literature Review on Immersive Virtual Reality in Education: State
  Of The Art and Perspectives}.
\newblock Technical report, 2015.

\bibitem{Graesser2004}
Arthur Graesser, Shulan Lu, G~Jackson, Heather Mitchell, Mathew Ventura, Andrew
  Olney, and Max Louwerse.
\newblock Autotutor: a tutor with dialogue in natural language.
\newblock 36:180--192, 06 2004.

\bibitem{Granitz2009NowStudents}
Neil~A. Granitz, Stephen~K. Koernig, and Katrin~R. Harich.
\newblock {Now It's Personal: Antecedents and Outcomes of Rapport Between
  Business Faculty and Their Students}.
\newblock {\em Journal of Marketing Education}, 31(1):52--65, 4 2009.

\bibitem{Gris2018VirtualAgents}
Ivan Gris and David Novick.
\newblock {Virtual Agent Interaction Framework (VAIF): A Tool for Rapid
  Development of Social Agents}.
\newblock In {\em AAMAS '18 Proceedings of the 17th International Conference on
  Autonomous Agents and MultiAgent Systems}, pages 2230--2232, Stockholm,
  Sweden, 2018.

\bibitem{Guenette2007IsWriting}
Danielle Gu{\'{e}}nette.
\newblock {Is feedback pedagogically correct?: Research design issues in
  studies of feedback on writing}.
\newblock {\em Journal of Second Language Writing}, 16(1):40--53, 3 2007.

\bibitem{Hartholt2013AllToolkit}
Arno Hartholt, David Traum, Stacy Marsella, Ari Shapiro, Giota Stratou, Anton
  Leuski, Louis-Philippe Morency, and Jonathan Gratch.
\newblock {All Together Now Introducing the Virtual Human Toolkit}.
\newblock In {\em International Workshop on Intelligent Virtual Agents}, pages
  368--381. Springer, Berlin, Heidelberg, 2013.

\bibitem{Huang2016}
Y.~Huang and M.~Kallmann.
\newblock Planning motions and placements for virtual demonstrators.
\newblock {\em IEEE Transactions on Visualization \& Computer Graphics},
  22(5):1568--1579, May 2016.

\bibitem{Kirk2006ComparingTasks}
David Kirk and Danae Stanton~Fraser.
\newblock {Comparing remote gesture technologies for supporting collaborative
  physical tasks}.
\newblock In {\em Proceedings of the SIGCHI conference on Human Factors in
  computing systems - CHI '06}, page 1191, New York, New York, USA, 2006. ACM
  Press.

\bibitem{Kulhavy1985FeedbackEfficiency}
Raymond~W. Kulhavy, Mary~T. White, Bruce~W. Topp, Ann~L. Chan, and James Adams.
\newblock {Feedback complexity and corrective efficiency}.
\newblock {\em Contemporary Educational Psychology}, 10(3):285--291, 7 1985.

\bibitem{Lester1997TheAgents}
James~C. Lester, Sharolyn~A. Converse, Susan~E. Kahler, S.~Todd Barlow,
  Brian~A. Stone, and Ravinder~S. Bhogal.
\newblock {The persona effect: affective impact of animated pedagogical
  agents}.
\newblock In {\em Proceedings of the SIGCHI conference on Human factors in
  computing systems - CHI '97}, pages 359--366, New York, New York, USA, 1997.
  ACM Press.

\bibitem{Nelson2009ThePerformance}
Melissa~M. Nelson and Christian~D. Schunn.
\newblock {The nature of feedback: how different types of peer feedback affect
  writing performance}.
\newblock {\em Instructional Science}, 37(4):375--401, 7 2009.

\bibitem{Pridemore1995ControlInstruction}
Doris~R. Pridemore and James~D. Klein.
\newblock {Control of Practice and Level of Feedback in Computer-Based
  Instruction}.
\newblock {\em Contemporary Educational Psychology}, 20(4):444--450, 10 1995.

\bibitem{Rickel1997}
Jeff Rickel and W.~Lewis Johnson.
\newblock Steve: An animated pedagogical agent for procedural training in
  virtual environments (extended abstract).
\newblock {\em SIGART Bulletin}, 8:16--21, 1997.

\bibitem{shang2019effects}
Xiumin Shang, Marcelo Kallmann, and Ahmed~Sabbir Arif.
\newblock Effects of correctness and suggestive feedback on learning with an
  autonomous virtual trainer.
\newblock In {\em Proceedings of the 24th International Conference on
  Intelligent User Interfaces: Companion}, pages 93--94, 2019.

\bibitem{Xiumin2019}
Xiumin Shang, Marcelo Kallmann, and Ahmed~Sabbir Arif.
\newblock Effects of virtual agent gender on user performance and preference in
  a vr training program.
\newblock In {\em Proceedings of the Future of Information and Communication
  Conference (FICC)}, 2019.

\bibitem{Shute2008FocusFeedback}
Valerie~J. Shute.
\newblock {Focus on Formative Feedback}.
\newblock {\em Review of Educational Research}, 78(1):153--189, 3 2008.

\bibitem{Sklar2003AgentsThing}
Elizabeth Sklar and {Elizabeth}.
\newblock {Agents for education: when too much intelligence is a bad thing}.
\newblock In {\em Proceedings of the second international joint conference on
  Autonomous agents and multiagent systems - AAMAS '03}, page 1118, New York,
  New York, USA, 2003. ACM Press.

\bibitem{Smith2018CommunicationReality}
Harrison~Jesse Smith and Michael Neff.
\newblock {Communication Behavior in Embodied Virtual Reality}.
\newblock In {\em Proceedings of the 2018 CHI Conference on Human Factors in
  Computing Systems - CHI '18}, pages 1--12, New York, New York, USA, 2018. ACM
  Press.

\bibitem{Tickle-Degnen1990TheCorrelates}
Linda Tickle-Degnen and Robert Rosenthal.
\newblock {The Nature of Rapport and Its Nonverbal Correlates}.
\newblock {\em Source: Psychological Inquiry}, 1(4):285--293, 1990.

\bibitem{VanderKleij2015EffectsOutcomes}
Fabienne~M. Van~der Kleij, Remco C.~W. Feskens, and Theo J. H.~M. Eggen.
\newblock {Effects of Feedback in a Computer-Based Learning Environment on
  Students’ Learning Outcomes}.
\newblock {\em Review of Educational Research}, 85(4):475--511, 12 2015.

\end{thebibliography}

\newpage
\onecolumn

\section*{\huge{Appendix}}

\vspace{1cm}

Table~\ref{tab:ssq} provides the quantitative data collected from the  Simulator Sickness Questionnaire (SSQ).



\vspace{1cm}

\begin{table}[h]
  \centering
  \begin{tabular}{l c c c c}
    {\textbf{ SSQ Symptoms}}
    &{\textbf{Nausea}}
    & {\textbf{Oculomotor}}
      & {\textbf{Disorientation}}
    & {\textbf{Total Score} }\\
    \midrule
    General discomfort& 38.16&	7.58&	0&	18.7 \\
    Fatigue	&9.54	&7.58	&0	&7.48 \\
    Headache	&19.08	&0	&0	&7.48\\
    Eye strain	&76.32	&7.58	&13.97	&37.4\\
    Difficulty focusing	&19.08	&22.74	&0	&18.7\\
    Salivation increasing	&9.54	&0	&0	&3.74\\
    Sweating	&0	&0	&0	&0\\
    Nausea	&9.54	&0	&0	&3.74\\
    Difficulty concentrating	&28.62	&0	&0	&11.22\\
    Fullness of the Head	&9.54	&0	&0	&3.74\\
    Blurred vision	&38.16	&7.58	&13.97	&22.44\\
    Dizziness with eyes open	&19.08	&7.58	&0	&11.22\\
    Dizziness with eyes closed	&19.08	&0	&0	&7.48\\
    Vertigo	&9.54	&0	&0	&3.74\\
    Stomach awareness	&0	&0	&0	&0\\
    Burping	&0	&0	&0	&0\\
    \midrule
    Mean	&19.08	&3.79	&1.75	&9.82\\
    SD	&18.78	&5.99	&4.62	&9.80\\

    \midrule
  \end{tabular}
  \caption{Simulator Sickness Questionnaire (SSQ) scale means.
  }~\label{tab:ssq}
\end{table}

\end{document}